\documentclass[prb,twocolumn]{revtex4-2}

\usepackage{graphicx}
\usepackage{dcolumn}
\usepackage{bm}
\usepackage{mathrsfs}
\usepackage{subfigure}
\usepackage{natbib}
\usepackage{amsmath}
\usepackage{amssymb}
\usepackage{Jonasmacros}
\usepackage{mathptmx}
\usepackage{txfonts}
\usepackage{ifsym}

\usepackage[colorlinks,citecolor=magenta,linkcolor=red]{hyperref}
\usepackage[usenames,dvipsnames,svgnames]{xcolor}

\newcommand{\jonas}[1]{\textcolor{black}{#1}}

\begin{document}
\date{\today}

\date{\today}

\title{Phonon mediated spin-spin interactions}

\author{J. Fransson}
\email{Jonas.Fransson@physics.uu.se}
\affiliation{Department of Physics and Astronomy, Box 516, 752 37, Uppsala University, Uppsala, Sweden}

\begin{abstract}
Indirect long range interactions between localized magnetic moments are in metals mediated by itinerant electrons. In insulators and semi-conductor, such interactions \jonas{cannot be mediated by itinerant electrons, however, delocalized electrons in the valence band may provide such a medium. Nevertheless, t}he existence of magnetically ordered insulators, for instance, metal-oxides, is therefore an everlasting source for proposals of various mechanisms that may support the order. Here, phonon mediated interactions between localized magnetic moments is considered as a mechanism that can provide quantifiable symmetric and anti-symmetric anisotropic spin-spin interactions. It is demonstrated that while a symmetric anisotropic interaction exists for all types of phonons, the existence of anti-symmetric anisotropic interactions requires broken inversion symmetry. The latter mechanism may explain the weak ferromagnetic order observed in chiral, e.g., CuO and CoO compounds. Furthermore, the interaction is nearly independent of the temperature at low temperature while approaches a linear growth at high. Spatially, the interactions have an oscillatory power law decay with the inter-\jonas{atomic} distance.
\end{abstract}
\maketitle

\section{Introduction}
Magnetic interactions between localized moments, can be categorized as direct and indirect interactions. While the former originates from overlapping states in the Coulomb integral, the latter may rather be seen as the result of more or less freely moving carriers that mediate the interactions between the moments. Especially, the latter type is often considered in metallic compounds where the natural carriers are the itinerant electrons, which move freely and, thereby, transferring the information about the moments across the structure \cite{PhysRev.96.99,ProgTheorPhys.16.45,PhysRev.106.893}. Moreover, despite the lack of freely moving carriers in insulators and semi-conductors, the exchange of electrons between magnetic and non-magnetic \jonas{atoms} gives rise to an effective interaction between the magnetic \jonas{atoms}, through the so-called super-exchange \cite{Physica.1.182,PhysRev.79.350,PhysRev.100.564,JPhysChemSolids.2.87,JPhysChemSolids.2.287} and double-exchange \cite{PhysRev.82.403,PhysRevLett.118.027203} mechanisms. Both these mechanisms may give rise to ferromagnetic, anti-ferromagnetic, and non-collinear, orders.

Recent discoveries of weak ferromagnetism in chiral metal-oxides \cite{JPhysChemC.123.3024,JPhysChemC.124.22610,ACSNano.16.12145} put the issue in a new perspective, since such compounds, e.g., CuO \cite{PhysSolidState.43.878,PhysRevB.64.174420,PhysSolidState.45.304}, NiO \cite{JApplPhys.70.6977}, CoO \cite{PhysRevB.6.4294}, are often anti-ferromagnetically ordered below some critical temperature and paramagnetic above. This is particularly intriguing, since the magnetic order is typically considered to be a result of the super-exchange mechanism. It was, therefore, speculated that the weak ferromagnetism may arise from uncompensated moments due to the chiral structure \cite{ACSNano.16.12145}.

While the interactions that stabilize magnetic orders are typically considered to be mediated by electrons, recent discoveries indicate the interactions may also be mediated by phonon, i.e., collective formation of nuclear motion. For instance, the increased coercive field with increasing temperature in metal-organic compounds \cite{JPhysChemLett.7.4988,ACSNano.14.16624,JPhysChemC.125.9875,JACS.142.17572} can be explained by introducing phonons which interact with the electrons and facilitate a stabilization of the magnetic order \cite{JPhysChemLett.14.2558,JPhysChemLett.14.5119}. Chiral phonons \cite{PhysRevLett.112.085503} have been suggested to generate spin-polarization among itinerant electrons \cite{NatMaterials.22.322,PhysRevResearch.5.L022039}, as well as stabilizing and enhancing the chiral induced spin selectivity effect \cite{PhysRevB.102.235416,JPhysChemC.126.3257}.

In this article, an expression for the phonon mediated spin-spin interaction is derived and given on the form of a Kubo-like formula. It is shown that this interaction comprises two anisotropic contributions, one symmetric and one anti-symmetric, however, there is no naturally emerging isotropic contribution. It is, moreover, shown that while the symmetric anisotropy exists for all types of phonons, the existence of the anti-symmetric anisotropy requires phonons with broken inversion symmetry, pertaining to chiral structures. The spatial decay and thermal properties of the interaction are discussed. In particular, it is shown that the thermal activation of the phonons leads to that the interaction may grow stronger with increasing temperature. Spatially, the interaction has an oscillatory power law decay, e.g., $1/r^{d-1}$ for acoustic phonons in $d$ dimension.

\section{Effective spin-phonon coupling}
The model introduced for the purpose pertains to general insulating and semi-conducting structures comprising localized magnetic moments $\bfM_m$, located at the spatial coordinate $\bfr_m$. The valence electrons are localized around the the nuclei, which justifies to model the local interaction between the electrons and $\bfM_m$ as $\sum_{ij}v_{ijm}\psi^\dagger_i(\bfr_m)\bfsigma\psi_j(\bfr_m)\cdot\bfM_m$, where $v_{ijm}$ is the direct exchange integral between the electrons in the orbirals $\phi_i(\bfr_m)$ and $\phi_i(\bfr_m)$ and the localized spin moment. Here, the spinor $\psi_i(\bfr_m)=(\psi_{i\up}(\bfr_m)\ \psi_{i\down}(\bfr_m))^t$ represents the annihilation of an electron in the orbital $i$, and $\bfsigma$ is the vector of Pauli matrices. Simultaneously, the electrons are connected to the nuclear displacement $\bfQ_m$ through $\sum_{ij}\psi^\dagger_i(\bfr_m)\psi_j(\bfr_m)\bflambda_{ijm}\cdot\bfQ_m$, where $\bflambda_{ijm}$ is the electron-phonon coupling.

\begin{figure}[t]
\begin{center}
\includegraphics[width=\columnwidth]{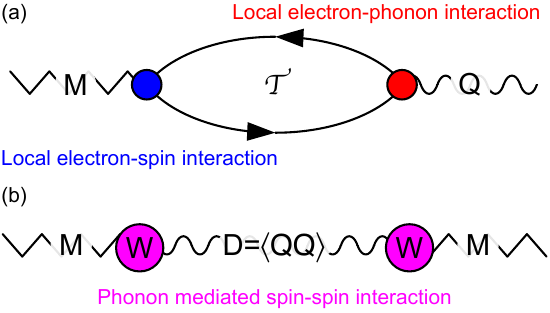}
\end{center}
\caption{(a) Spin-phonon interaction at a nucleus mediated by the local valence electrons in local (blue dot) spin-electron and (red dot) electron-phonon interactions, and (b) effective phonon mediated spin-spin interaction between \jonas{atoms}. Spin, electron, and phonon are indicated by the zig-zag, arrowed, and wavy lines, respectively.}
\label{fig-SpinPhonon}
\end{figure}

Using the results derived in Ref. \cite{PhysRevMaterials.1.074404}, the coupling between the spin and displacement operators at the coordinate $\bfr_m$ can be formulated through the two-components
\begin{subequations}
\begin{align}
\bfM_m\cdot\calT^\text{sc}_m\cdot\bfQ_m
	,
\\
\bfQ_m\cdot\calT^\text{cs}_m\cdot\bfM_m
	,
\end{align}
\end{subequations}
which are schematically illustrated in Fig. \ref{fig-SpinPhonon} (a).
The coupling $\calT_m^\text{sc}=\calT^\text{sc}(\bfr_m)$ is defined as a spin-charge susceptibility, represented by the loop in Fig. \ref{fig-SpinPhonon} (a), and in general given by the time-dependent expression
\begin{align}
\calT^\text{sc}_m(t,t')=&
	2\sum_{ijkl}
		v_{ijm}
		\eqgr{\bfs_{ij}(\bfr_m,t)}{n_{kl}(\bfr_m,t')}\bflambda_{klm}
		,
\label{eq-Tsc}
\end{align}
where $\bfs_{ij}=\psi^\dagger_i\bfsigma\psi_j/2$ and $n_{kl}=\psi^\dagger_k\psi_l$ represent the electron spin and charge operators, respectively. It can be noticed that the coupling $\calT^\text{cs}_m$ is obtained by interchanging the order of $v_{ijm}\bfs_{ij}$ and $n_{kl}\bflambda_{klm}$ in Eq. \eqref{eq-Tsc}.
The coupling tensor $\calT^\text{sc}_m$ is, hence, representing the electrons in the orbits around the nucleus and is thereby connecting the localized spin moment with the nuclear motion. 

These aspects of the model can, then, be summarized as the Hamiltonian
\begin{align}
\Hamil=&
	\Hamil_\text{ph}
	-
	\sum_m
		\Bigr(
			\gamma\bfM_m\cdot\bfB_\text{ext}
			+
			\bfM_m\cdot\calT^{sc}_m\cdot\bfQ_m
			+
			\bfQ_m\cdot\calT^{cs}_m\cdot\bfM_m
		\Bigr)
	,
\end{align}
where the first term, $\Hamil_\text{ph}$, represents the phonon reservoir and where external magnetic field $\bfB_\text{ext}$ has been included, with $\gamma=g\mu_B$, where $g$ and $\mu_B$ is the gyromagnetic ratio and Bohr magneton, respectively.

The two coupling tensors $\calT^\text{sc}_m$ and $\calT^\text{cs}_m$ are connected through the transpose as $\{\calT^\text{sc}_m\}^t=\calT^\text{cs}_m$. It should, moreover, be noticed that the tensor $\calT^\text{sc}_m$ is non-vanishing whenever the electrons captured in the tensor are either spin-polarized and/or are mixed via spin-orbit coupling \cite{PhysRevMaterials.1.074404}. Since spin-orbit coupling is non-vanishing for all elements but H, it is safe to assume that the coupling tensor in the pertinent structures is non-zero. This establishes a connection between the localized spin moment and the nuclear motion.

In preparation for the next step, the displacement operator is expressed in terms of the phonon destruction and creation operators $b_p$ and $b^\dagger_p$, using the plane wave expansion $\bfQ_m(t)=\sum_\mu\int l_q\bfepsilon_qQ_q(t)e^{i\bfq\cdot\bfr_m}d\bfq/\Omega$, where $\Omega$ is the associated volume. The subscript $q=(\bfq,\mu)$ denotes the wave vector $\bfq$ and branch $\mu$ associated with the displacement operator $Q_q=b_q+b^\dagger_{\bar{q}}$ ($\bar{q}=(\bar\bfq,\mu)$, $\bar\bfq=-\bfq$) and phonon energy $\omega_q$. The quantity $l_q=\sqrt{\hslash/2\omega_q\rho\Omega}$ defines a length scale in terms of the density $\rho$, and $\bfepsilon_q$ is the displacement polarization vector. Here, it can be observed that $\bfepsilon_q^*=\bfepsilon_{\bar{q}}\neq\bfepsilon_q$, in general, while the equality $\bfepsilon_{\bar{q}}=\bfepsilon_q$ holds in inversion symmetric structures.

\section{Phonon mediated spin-spin interactions}
The expression for the spin interactions is derived using the approach in Refs. \cite{PhysRevB.82.180411,PhysRevMaterials.1.074404}. This approach enables an order-by-order expansion of the partition function $\calZ={\rm Tr}\exp\{i\int(i\dt-\Hamil)dt\}$, where the trace ${\rm Tr}$ runs over the pertinent degrees of freedom. In the present context, the focus lies on a bilinear form for the spin moment operators $\bfM$, see Fig. \ref{fig-SpinPhonon} (b), which can be formulated as ($x=(\bfr,t)$)
\begin{subequations}
\label{eq-MDM}
\begin{align}
\delta\calS_\text{M}=&
	-\frac{1}{2}
	\int
		\bfM(x)\cdot\calD(x,x')\cdot\bfM(x')
	dxdx'
	,
\label{eq-dS}
\\
\calD(x,x')=&
	\int
		W_p(x,\tau)
		D_{pq}(\tau,\tau')
		W_q^\dagger(\tau',x')
\nonumber
\\&\times
		e^{i\bfp\cdot\bfr-i\bfq\cdot\bfr'}
	\frac{dp}{\Omega}
	\frac{dq}{\Omega}
	d\tau
	d\tau'
	.
\label{eq-calD}
\end{align}
\end{subequations}
In this expression, $W_p(x,\tau)=l_p\calT_\bfr^{sc}(t,\tau)\cdot\bfepsilon_q$ denotes the effective spin-phonon coupling, $D_{qq'}(t,t')=\eqgr{Q_q(t)}{Q_{\bar{q}'}(t)}$ is the phonon propagator, whereas integration over $p$ and $q$ include both summation over the modes and wave vectors.

Formulated as in Eq. \eqref{eq-MDM}, the different parts of the interaction are distinguishable. The magnetic moment $\bfM(x)$ interacts via exchange with the electrons in its vicinity around the nucleus. These electrons are also coupled to the phonons, and the propagation of the phonons ties the magnetic moments $\bfM(x)$ and $\bfM(x')$ together. While the properties of the electrons are discussed above, the next task is to consider how the quality of the phonons influences the resulting properties of the spin-spin interactions.

The symmetry structure of the interaction tensor $\calD(x,x')$ is clarified by introducing the symmetric and anti-symmetric components $\mathbb{S}(x,x')=\{\calD(x,x')+[\calD(x,x')]^t\}/2$ and $\mathbb{A}(x,x')=\{\calD(x,x')-[\calD(x,x')]^t\}/2$, respectively, such that $\calD(x,x')=\mathbb{S}(x,x')+\mathbb{A}(x,x')$. In terms of the subscripts $\alpha\beta$, referring to the matrix row $\alpha$ and column $\beta$, respectively, the symmetric tensor has the property
\begin{align}
\mathbb{S}_{\alpha\beta}(x,x')=&
	\frac{1}{2}
	\Bigl(
		\calD_{\alpha\beta}(x,x')
		+
		\calD_{\beta\alpha}(x',x)
	\Bigr)
	.
\end{align}
This implies that $\mathbb{S}_{\alpha\alpha}(x,x)=\calD_{\alpha\alpha}(x,x)$ and $\mathbb{S}_{\beta\alpha}(x',x)=\mathbb{S}_{\alpha\beta}(x,x')$, as expected for a symmetric tensor. However, in stark contrast to the electronically mediated spin-spin interactions, there is no natural definition of an isotropic interaction of Heisenberg type. It is, nonetheless, possible to define an isotropic interaction parameter, for instance, by setting $J(x,x')\propto\tr\mathbb{S}(x,x')$. On the other hand, this is an artificial construction only making sense under the restricted conditions that $\mathbb{S}_{\alpha\beta}=\delta_{\alpha\beta}\mathbb{S}$. Therefore, defining an isotropic interaction has little value since the symmetric anisotropy tensor cannot be discarded on the basis of symmetry arguments.

The symmetric interaction tensor is non-vanishing under very general conditions, which can be seen from the following observation. The property of $\calT^{sc}_\bfr$ under the transpose leads to that the anti-symmetric tensor can be written
\begin{align}
\mathbb{S}(x,x')=&
	\frac{1}{2}
	\int
		\biggl(
			D_{pq}(\tau,\tau')e^{i\bfp\cdot\bfr-i\bfq\cdot\bfr'}
			+
			D_{qp}(\tau,\tau')e^{i\bfq\cdot\bfr-i\bfp\cdot\bfr'}
		\biggr)
\nonumber
\\&\times
		W_p(x,\tau)W_q^\dagger(\tau',x')
	\frac{dp}{\Omega}
	\frac{dq}{\Omega}
	d\tau
	d\tau'
	.
\label{eq-S}
\end{align}
Further insight to the phononic impact on this interaction is be achieved by assuming non-interacting and time-independent phonons, such that $D_{pq}(t,t')=\delta_{pq}d_q(t-t')$. Here, $d_q(t-t')$ is the unperturbed phonon propagator. The symmetric tensor can, then, be Fourier transformed and written as
\begin{align}
\mathbb{S}(\bfr,\bfr';z)=&
		\int
			W_q(\bfr)W_{\bar{q}}^\dagger(\bfr')
			d_q(z)
			\cos\bfq\cdot\bfR
		\frac{dq}{\Omega}
	,
\label{eq-Srr}	
\end{align}
where $\bfR=\bfr-\bfr'$ and $W_p(\bfr)=\int W_p(x,\tau)e^{-i\omega(t-\tau)}d\tau$. The consequence of this expression has the far-reaching implication that there must exist a phonon mediated spin-spin interaction regardless of the crystal symmetry, hence, also for inversion symmetric structures in which $\bfepsilon_{\bar{q}}=\bfepsilon_q$. This conclusion can be understood by the following further simplifications. Assuming that $l_q$ and $\bfepsilon_q$ are slowly varying functions of the wave vector $\bfq$, and that the phonon propagator $d_q$ depends on the length of $\bfq$, leads to that the interaction can be written
\begin{align}
\mathbb{S}(\bfr,\bfr';z)=&
	\frac{1}{\pi}
	\sum_\mu
		W_\mu(\bfr)W_\mu^\dagger(\bfr')
		\int_0^{q_c}
			d_q(z)j_0(qR)
		\frac{q^2dq}{2\pi}
\label{eq-Sinversion}
	,
\end{align}
where $j_0(x)=\sin(x)/x$ is the zeroth order spherical Bessel function, $q_c$ is the phonon high energy cut-off, and $R=|\bfR|$. Since the integral over $q$ is non-zero it follows that the symmetric interaction is non-vanishing.

The conclusion that can be drawn from the above arguments is that the symmetric interaction is non-vanishing under rather general conditions. This is clear from the observation that $\mathbb{S}$ remains non-zero under the severe restrictions imposed on the expression in Eq. \eqref{eq-Sinversion}. Then, in a more general structure with less symmetry restrictions, also the integrand in Eq. \eqref{eq-S} comprises less compensating contributions that could lead to a vanishing interaction.

In perspective, another source of spin-phonon coupling is provided through the vibrationally induced spin-orbit coupling \cite{PhysRevResearch.5.L022039}. This mechanism is fundamentally different from the one discussed here, since the vibrationally assisted spin-orbit coupling exists only in structures with broken inversion symmetry.
\jonas{
This can be directly seen by a plane wave expansion of the spin-orbit coupling, yielding the electron-phonon component \cite{PhysRevResearch.5.L022039}
\begin{align}
\Hamil_\text{SOC}^{(1)}=&
	-\xi
	\int
		l_q\bfepsilon_q\cdot\bfq V(\bfq)
		\psi^\dagger_{\bfk+\bfq}(\bfk\times\bfq)\cdot\bfsigma\psi_\bfk Q_q
	\frac{dq}{\Omega}
	\frac{d\bfk}{\Omega}
	.
\end{align}
Obviously, for structures with broken inversion symmetry, the polarization vector $\bfepsilon_{\bar{q}}\neq\bfepsilon_q$, which leads to that the Hamiltonian is non-vanishing, in general.
}

By contrast, for the anti-symmetric tensor to be non-vanishing, the inversion symmetry has to be broken, which can be understood by the following argument. First, it should be noticed that the anti-symmetric tensor $\mathbb{A}$ has the property $\mathbb{A}_{\alpha\beta}(x,x')=[\calD_{\alpha\beta}(x,x')-\calD_{\beta\alpha}(x',x)]/2$. Hence, $\mathbb{A}_{\alpha\alpha}(x,x)=0$ and $\mathbb{A}_{\beta\alpha}(x',x)=-\mathbb{A}_{\alpha\beta}(x,x')$.
A direct consequence of these properties is that it enables defining the product
\begin{align}
\bfD(x,x')\cdot
	\Bigl(\bfM(x)\times\bfM(x')\Bigr)
	,
\end{align}
where $\bfD(x,x')=[\mathbb{A}_{yz}(x,x'),\mathbb{A}_{zx}(x,x'),\mathbb{A}_{xy}(x,x')]$ is a three component vector. The anti-symmetric tensor can, thus, be reformulated as a Dzyaloshinskii-Moriya\jonas{-like} interaction.

The conditions that have to be fulfilled for $\bfD$ to be non-vanishing is traced back to the definition of the interaction tensor in Eq. \eqref{eq-calD}. Repeating the analysis that was made for $\mathbb{S}$ and reducing to the unperturbed phonon propagator, leads to that
\begin{align}
\mathbb{A}(\bfr,\bfr';z)=&
	i
	\int
		W_q(\bfr)W_{\bar{q}}^\dagger(\bfr')
		d_q(z)
		\sin\bfq\cdot\bfR
	\frac{dq}{\Omega}
	.
\label{eq-A}
\end{align}
This expressions clearly shows that the properties of the interaction are governed by the symmetry conditions of the phonons. Especially, the presence of the function $\sin\bfq\cdot\bfR$, which is odd under inversion symmetry operations, explicitly indicates that the polarization vectors and/or the phonon energy have to posses additional structure for the anti-symmetric interaction to be non-vanishing.

The most conspicuous and important requirement is that the inversion symmetry has to be broken. If not, the polarization vectors as well as the phonon propagator are inversion symmetric. Hence, the anti-symmetric inversion property of $\sin\bfq\cdot\bfR$ leads to that the integral in Eq. \eqref{eq-A} vanishes.

Here, a comparison with electronically mediated spin-spin interactions is instructive. It is known that in absence of an inversion symmetry breaking mechanism, manifested by, for instance, spin-orbit coupling or scatterers that generate spin currents, spin-degenerate electrons cannot give rise to a Dzyaloshinskii-Moriya interaction \cite{PhysRevB.108.224408}. The preservation of inversion symmetry can for phonons be regarded to have a similar effect.

On the contrary, should the inversion symmetry in the structure be broken, the polarization vector $\bfepsilon_q$ has a non-trivial angular dependence which changes the situation significantly. The polarization vector is angular dependent whenever the space that spans the phononic structure is inhomogeneous, since the inhomogeneity introduces a directional dependence of the momentum variable $\bfq$. The angular dependence can also be related to that the polarization coordinates of $\bfepsilon_q$ cannot be separated from one another.

An example of such a dependence pertaining to helical, hence, chiral phonons is provided by the polarization vector $\bfepsilon_\bfq=(a\cos\varphi,a\sin\varphi,\mathcent\varphi)/d(\varphi)$, where $a$ and $c=2\pi\mathcent$ defines the radius and pitch, respectively, of the helical structure in reciprocal space whereas $d^2(\varphi)=a^2+\mathcent^2\varphi^2$. The longitudinal coordinate $(\bfepsilon_\bfq)_z=\mathcent\varphi/d(\varphi)$ directly depends on the rotation angle $\varphi$. Then, for free phonons in the cylindrical space ($(\bfq_\perp,q_\|)$ and $(\bfR_\perp,R_\|)$), by integrating out the longitudinal coordinate $q_\|$, $0\leq q_\|\leq q_L\varphi/2\pi$, one obtains for a single mode
\begin{align}
\mathbb{A}(\bfr,\bfr';z)=&
	\int_{\calS_\perp}
		W_\bfq(\bfr)W_{\bar\bfq}^\dagger(\bfr')
		d_\bfq(z)
	\frac{\calR_\varphi}{R_\|}
		\cos\Bigl(\bfq_\perp\cdot\bfR_\perp-\delta_\varphi\Bigr)
	\frac{d\bfq_\perp}{\Omega}
	,
\label{eq-Asimplified}
\end{align}
where $\calR_\varphi=|\sin(R_\|q_L\varphi/4\pi)|$ and $\tan\delta_\varphi=\cot(R_\|q_L\varphi/4\pi)$, whereas $\calS_\perp$ is the surface over which the transverse variables are integrated. The remaining integrand is neither even nor odd as function of the azimuthal angle $\varphi$. Hence, the resulting anti-symmetric interaction is under such conditions non-vanishing, in general.

It is clear from this discussion that the anti-symmetric interaction requires broken inversion symmetry to be non-vanishing. Therefore, the anti-symmetric interaction is less generally occurring than the symmetric, since the latter tends to be preserved also in highly symmetric structures.

\section{Thermal and spatial properties of the interaction tensor}
Because of the phononic mediation, the spin-spin interactions are expected to acquire a non-trivial temperature dependence. Essentially, the interaction strength is expected to increase with increasing temperature, a property that is easily be corroborated by considering the Hamiltonian $\Hamil_\text{M}$ corresponding to the action variable $\delta\calS_\text{M}$, which is obtained by setting the times $t'=t$. The resulting spin-spin interaction, Eq. \eqref{eq-calD}, when converted to real time-variables assumes the form 
\begin{widetext}
\begin{align}
\calD(\bfr,\bfr';t)=&
	\int\theta(t-\tau)
		l_pl_q
		e^{i\bfp\cdot\bfr-i\bfq\cdot\bfr'}
		\Biggl(
			\calT_\bfr^{sc,>}(t,\tau)\cdot\bfepsilon_p
			\biggl[
				D_{pq}^r(\tau,\tau')
				\bfepsilon_q\cdot\calT_{\bfr'}^{cs,<}(\tau',t)
				+
				D_{pq}^<(\tau,\tau')
				\bfepsilon_q\cdot\calT_{\bfr'}^{cs,a}(\tau',t)
			\biggr]
\nonumber\\&
			-
			\calT_\bfr^{sc,<}(t,\tau)\cdot\bfepsilon_p
			\biggl[
				D_{pq}^r(\tau,\tau')
				\bfepsilon_q\cdot\calT_{\bfr'}^{cs,>}(\tau',t)
				+
				D_{pq}^>(\tau,\tau')
				\bfepsilon_q\cdot\calT_{\bfr'}^{cs,a}(\tau',t)
			\biggr]
		\Biggr)
	\frac{dp}{\Omega}
	\frac{dp}{\Omega}
	d\tau
	d\tau'
	.
\end{align}
\end{widetext}
In this expression, the superscripts $>$/$<$/$r$/$a$ refer to the lesser/greater/retarded/advanced forms of the propagators, whereas $\theta(t)$ is the Heaviside step function.

On the one hand, the electronic propagators $\calT_\bfr^{>/</a}$ pertain to the valence and conduction electrons in the metal oxide. For sufficiently low temperatures thermal excitations across the gap between the valence and conduction states are not supported; typically this gap is larger than 0.5 eV ($\sim$ 500 K). Hence, as the occupation of these states remain nearly constant, the electronic propagators can safely be assumed to not vary substantially for a wide range of temperatures.

On the other hand, the phonon occupation changes dramatically in this temperature range. The phonon occupation is captured in the phonon propagators $D_{pq}^{>/<}$ which account for the correlations between phonons characterized by $p$ and $q$. For non-interacting phonons, the lesser/greater Green function is given by
\begin{align}
d_q^{</>}(t,t')=&
	(-i)
	\biggl(
		n_B(\omega_q)
		e^{\mp i\omega_q(t-t')}
		+
		n_B(-\omega_q)
		e^{\pm i\omega_q(t-t')}
	\biggr)
	,
\end{align}
where $n_B(\omega)$ is the Bose-Einstein distribution function. In the stationary regime, both integrals $\int d_q^{</>}(t,t')dt'=\coth(\beta\omega_q/2)$, where $\beta=1/k_BT$ is the reciprocal temperature in terms of the Boltzmann constant $k_B$ and temperature $T$. Consequently, the phonon mediated spin-spin interaction is the weakest for low temperatures, since $\coth(\beta\omega_q/2)\rightarrow1$, $T\rightarrow0$. This should be contrasted with the asymptotically linear growth of the occupation ($\coth(\beta\omega_q/2)\sim T$, for large $T$), leading to an increasing interaction strength with increasing temperature.

Finally, the spatial decay of the interactions between two spin moments can be estimated by using the expressions in Eqs. \eqref{eq-Srr} and \eqref{eq-A}. Considering structures with broken inversion symmetry, the arguments of the trigonometric functions can be replaced by $(\bfq+\bfq_0)\cdot\bfR$, for some wave vector $\bfq_0$. In the limit of static interactions, $\lim_{z\rightarrow0}d_q(z)=-1/\omega_q$, and assuming that the effective electron-phonon coupling $W_q(\bfr)$ varies slowly with $q$, the symmetric interaction tensor becomes for acoustic phonons ($\omega_q=\gamma_\mu q$)
\begin{align}
\mathbb{S}(\bfr,\bfr')=&
	\left\{
	\begin{array}{lr}
	-\frac{2}{\pi^2}
	\sum_\mu W_\mu(\bfr)W_\mu^\dagger(\bfr')
		\frac{\cos\bfq_0\cdot\bfR}{\gamma_\mu R^2}
		\sin^2\frac{q_cR}{2}
			&
			(3D)
		\\\\
	-\frac{1}{2\pi^2}
	\sum_\mu W_\mu(\bfr)W_\mu^\dagger(\bfr')
		\frac{\cos\bfq_0\cdot\bfR}{\gamma_\mu R}
		I_0(R)
			&
			(2D)
	\end{array}
	\right.
\label{eq-spatialS}
\end{align}
where $I_0(R)=\int_0^{q_cR}J_0(x)dx\rightarrow1$, $q_cR\rightarrow\infty$, while the anti-symmetric interactions are given by
\begin{align}
\mathbb{A}(\bfr,\bfr')=&
	\left\{
	\begin{array}{lr}
		-\frac{2}{\pi^2}
		\sum_\mu W_\mu(\bfr)W_\mu^\dagger(\bfr')
			\frac{\sin\bfq_0\cdot\bfR}{\gamma_\mu R^2}
			\sin^2\frac{q_cR}{2}
			&
			(3D)
		\\\\
		-\frac{1}{2\pi^2}
		\sum_\mu W_\mu(\bfr)W_\mu^\dagger(\bfr')
			\frac{\sin\bfq_0\cdot\bfR}{\gamma_\mu R}
			I_0(R)
			&
			(2D)
	\end{array}
	\right.
\label{eq-spatialA}
\end{align}

\begin{figure}[t]
\begin{center}
\includegraphics[width=\columnwidth]{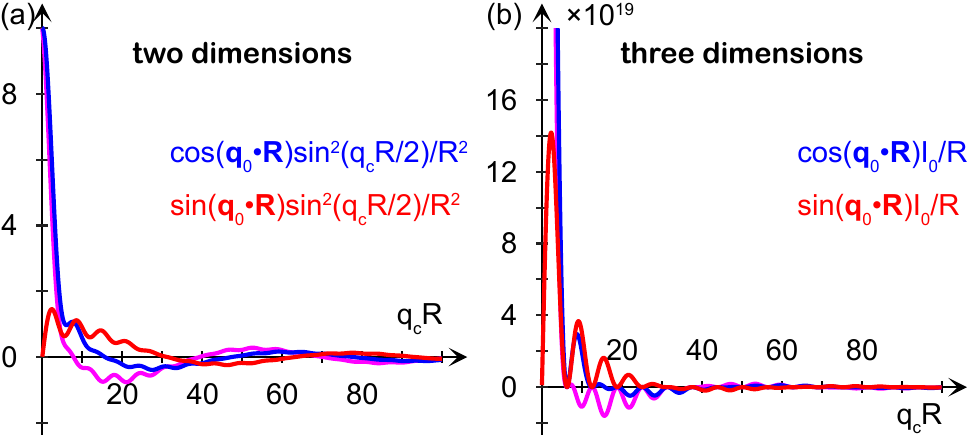}
\end{center}
\caption{Spatial decay of the (blue) symmetric and (red) anti-symmetric spin-spin interactions as given in Eqs. \eqref{eq-spatialS} and \eqref{eq-spatialA} for (a) two and (b) three dimensions, and (magenta) difference between the symmetric and anti-symmetric interactions. Here, $q_c=\pi/5$ \AA$^{-1}$ and $q_0=q_c/10$.}
\label{fig-spatial}
\end{figure}

The spin-spin interaction has a power law dependence on the distance and is, hence, long-range, as can be seen in Fig. \ref{fig-spatial} for (a) two and (b) three dimensions. Although both the symmetric (blue) and anti-symmetric (red) interactions have the same decay rate, their individual oscillatory dependence differ whereby the interactions are dominated by either the symmetric or anti-symmetric contribution depending on the distance between the spins.

Furthermore, for optical phonons, the spatially decaying functions $\sin^2(q_cR/2)/R^2$ and $I_0(R)/R$ are replaced by $\int_0^{q_cR}j_0(x)dx/R$ and $\int_0^{q_cR}J_0(x)x^{-1}dx$, respectively. The interaction mediated by optical phonons is, thus, further long-range.

It should be noticed that the symmetric interaction $\mathbb{S}$ changes between positive and negative signs with increasing distance inter-\jonas{atomic} distance $R$ whenever the inversion symmetry is broken, $\bfq_0\neq0$. Hence, while an inversion symmetric crystal ($\bfq_0=0$) may only assume anti-ferromagnetic ordering through this mechanism, the symmetric interaction may favor both ferro- and anti-ferromagnetic ordering in structures with no inversion symmetry.

\begin{figure}[t]
\begin{center}
\includegraphics[width=\columnwidth]{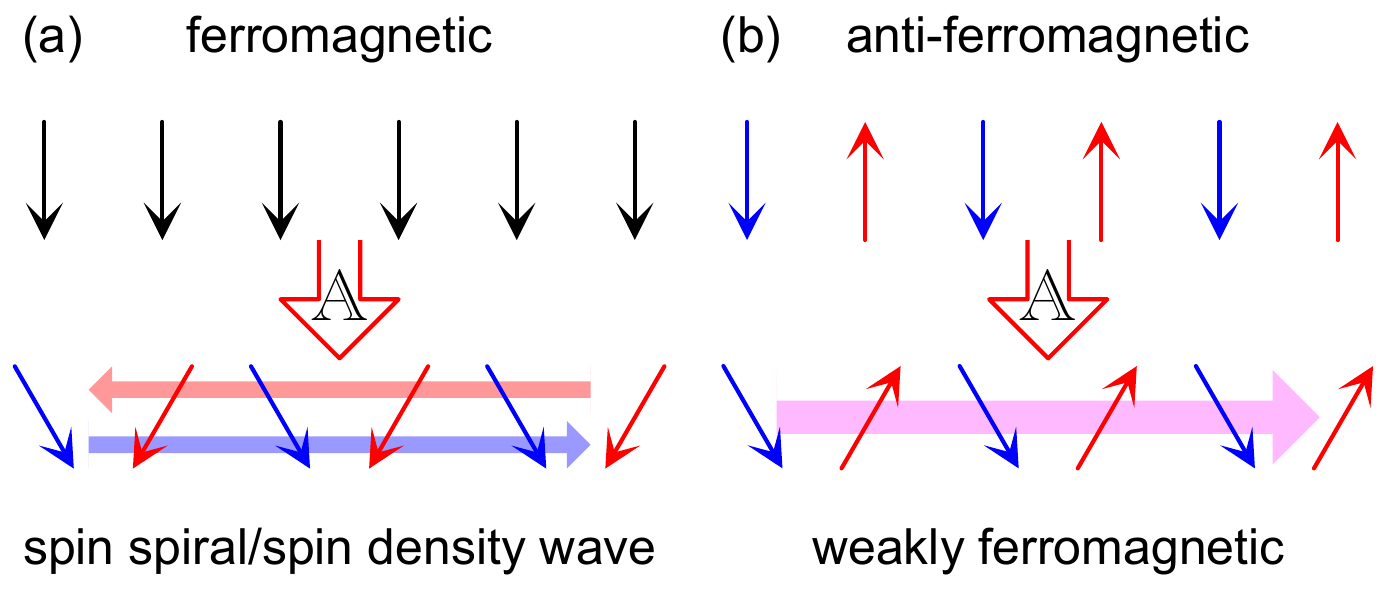}
\end{center}
\caption{Collinear (a) ferro- and (b) anti-ferromagnetic order is redefined to non-collinear \jonas{(a) spin-spiral or spin-density wave and (b) weakly} ferromagnetic order.}
\label{fig-CollvsNoncoll}
\end{figure}

The finiteness of the anti-symmetric interaction $\mathbb{A}$ has the effect to reorient the local magnetic moments from strictly collinear configurations governed by the symmetric interaction to a non-collinear magnetic texture. This emerging texture does, however, not randomize the spins but rather redefines the magnetic order by generating a finite component perpendicular to the collinear order, see Fig. \ref{fig-CollvsNoncoll}. It should be stressed, nevertheless, that a \jonas{while a} finite magnetic moment emerges in the weakly ferromagnetic order, the emergent \jonas{spin-spiral order or spin-density wave} modifies the overall ferromagnetic order without necessarily \jonas{completely} destroying it.

\jonas{
For the sake of comparison, consider the electronically mediated indirect spin-spin interactions provided from, e.g., Ref. \cite{PhysRevMaterials.1.074404}, which is calculated using the same formalism as here. An estimate of its spatial dependence of the isotropic, Heisenberg-like, interaction $J_{ij}$ is in three dimensions given by
\begin{align}
J_{ij}\sim&
	-\frac{v_iv_j}{4}
	\biggl(
		\frac{N_0}{\pi}
	\biggr)^3
	\dote{V}
	j_0(\kappa_Vr_i)y_0(\kappa_Vr_j)
	,
\label{eq-J}
\end{align}
where $N_0=m_e/\hslash^2\ [\sim2]$, $m_e$ is the electron mass, $\dote{V}$ is the valence band edge energy, $\kappa_V=\sqrt{2N_0\dote{V}}$, and $y_0(x)$ is the zeroth order spherical Neumann function.
}

\jonas{
The effective spin-phonon coupling $W_\mu$ can be estimated by roughly $v_i\bflambda_j\cdot\bfepsilon_\mu l_\mu=v_i\tilde\lambda_{j\mu}$ ($\tilde\lambda_{j\mu}=\bflambda_j\cdot\bfepsilon_\mu l_\mu$ effective electron-phonon coupling), such that the spatial dependence of the symmetric interaction in three dimensions assumes the form
\begin{align}
\mathbb{S}_{ij}\sim&
	-\frac{2}{\pi^2}
	v_i^2\tilde\lambda_{j\mu}^2
	\frac{\cos\bfq_0\cdot\bfR}{\gamma_\mu R}
	\sin^2\frac{q_cR}{2}
	.
\label{eq-Sest}
\end{align}
}

\jonas{
From this analysis one can conclude that nominally the phonon mediated spin interaction is expected to be smaller than the corresponding electronically mediated interaction between spin-moments near to each other. This is clear since while the latter scales as the square of the local exchange interaction $v_i$ (Eq. \ref{eq-J}), the former also scales as the square of the local effective electron-phonon interaction $\tilde\lambda_{i\mu}$ (Eq. \ref{eq-Sest}). Nonetheless, the range functions for electrons and phonons are significantly distinct. The former has a oscillatory decay that scales as $1/R^d$, where is the distance between the spins and $d$ the spatial dimension, while the latter has an oscillatory decay as $1/R^{d-1}$. Hence, phonons mediate interactions at a longer range than electrons do and may, thereby, govern the formation of the acquired long range order.
}

\section{Summary and discussion}
In conclusion, it has been analytically shown that local phonon mediated spin-spin interactions are generally present in insulating matter. The interactions build from the local coupling between valence electrons and localized spin moment, on the one hand, and between valence electrons and nuclear motion, on the other, according to the scheme derived and discussed in Ref. \cite{PhysRevMaterials.1.074404}. By coherent nuclear vibrations, phonons, these interactions are mediated between the local spin moments which, in general, are non-vanishing.

It was, moreover, shown that the interaction tensor, the phonon-mediated magnetic susceptibility, can be decomposed into symmetric and anti-symmetric contributions, of which the former is non-vanishing for general spatial symmetries of the compounds. The latter contribution, by contrast, requires a broken inversion symmetry to be non-vanishing. Hence, inversion symmetric structures are expected to stabilize collinear anti-ferromagnetic order which should be put in contrast with structures in which the inversion symmetry is broken. In a broken inversion symmetry compound, the symmetric interaction may also favor collinear ferromagnetic order. Furthermore, under such conditions the anti-symmetric contribution cants the localized spin moments in a direction perpendicular to the collinear orientation governed by the symmetric tensor contribution. Hence, a ferromagnetically ordered compound may become a \jonas{spin-spiral or spin-density wave} ordered one, whereas an anti-ferromagnetic may turn into a weak ferromagnet. This property may \jonas{provide novel and complementary ways to} explain the observations of weak ferromagnetism in\jonas{, e.g.,} chiral CuO \cite{ACSNano.16.12145}.

The phonon mediated spin-spin interactions are expected to grow with increasing temperature within a range of a few hundreds, up to a few thousands, of Kelvin, depending on the gap between the valence and conduction electron states. \jonas{Certainly, this estimate is obtained by disregarding other mechanisms that may play a crucial role at elevated temperatures, such as structural phase transitions among others. Despite this restriction it can be understood that t}he interactions are predicted to be weakest and nearly independent of the temperature at low temperatures, while it approaches a linear growth at high. It was finally shown that the interactions decay oscillatory with the inter-nuclei distance as a power law; $1/r^{d-1}$ for acoustic phonons in $d$ dimensions. Therefore, while the symmetric interaction tensor provides only anti-ferromagnetic ordering in structures with preserved inversion symmetry, the oscillatory spatial dependence in structures without inversion symmetry open up for both ferro- and anti-ferromagnetic ordering.

More experiments on, e.g, insulating chiral magnetic compounds as well as \emph{ab initio} simulations may shed further light on the properties of phonon mediated spin-spin interactions. Ways to test the predictions made in this paper include, for instance, studying the distance dependence between localized magnetic moment using magnetic molecules adsorbed on insulating substrate material.

\acknowledgements
The author thanks R. Naaman, L. Nordstr\"om, and D. Waldeck for helpful and inspiring discussion. A special thank is given to R. Naamam and D. Waldeck who suggested the research and provided experimental data. Funding from Olle Engkvists Stiftelse is acknowledged.

\bibliography{CISSref}

\end{document}